\RequirePackage{fix-cm}
\documentclass[twocolumn,epjc3]{svjour3}  
\smartqed  
\RequirePackage{graphicx}
\journalname{Eur. Phys. J. C}
\begin{document}

\title{Relativistic Landau levels for a fermion-antifermion pair interacting through Dirac oscillator interaction
}

\author{Abdullah Guvendi\thanksref{e1,addr1}}
\thankstext{e1}{e-mail: abdullah.guvendi@ksbu.edu.tr}

\institute{Medical Imaging Techniques, Simav Vocational School of Health Services, Kutahya Health Sciences University, TR- 43500 Kutahya, Turkey\label{addr1} }

\date{Received: date / Accepted: date}

\maketitle

\begin{abstract}
We introduce a unique model for a fermion antifermion pair interacting through Dirac oscillator interaction in the presence of external uniform magnetic field. In order to acquire a non perturbative energy spectrum for such a system we solve the corresponding form of a fully covariant two body Dirac equation. The dynamic symmetry of the system allows to study in three dimensions and the corresponding equation leads a $4\times4$ dimensional matrix equation for such a static and spinless composite system. We can obtain an exact solution of the matrix equation and arrive at a spectrum (in closed form) in energy domain. The obtained energy spectrum shows that the composite system that under scrutiny behaves like a single relativistic quantum oscillator. As a result, we obtain relativistic Landau levels of a fermion antifermion pair interacting through Dirac oscillator interaction and determine the components of the corresponding bi-spinor. We think that our results can provide enlightening informations about the quark antiquark systems and thus mass formula for mesons.
\end{abstract}

\section{Introduction}
\label{intro}

\qquad The Dirac oscillator has been introduced as a new type of interaction in the Dirac equation \cite{ito1967example,cook1971relativistic,moshinsky1989dirac} (also see \cite{nikolsky1930oszillatorproblem}). The corresponding Hamiltonian to this equation is linear in both momentum and spatial coordinate \cite{ito1967example,cook1971relativistic} and this instance of Dirac equation gives a simple quantum oscillator solution with a strong spin-orbit coupling term in the non-relativistic limit \cite{moshinsky1989dirac,carvalho2011dirac}. This is why it has been called as Dirac oscillator \cite{moshinsky1989dirac,book:1028629}. It has been shown that the Dirac oscillator interaction corresponds to an anomalous magnetic interaction when this interaction is written in the manifestly covariant form in the Dirac equation \cite{bentez1990solution}. Taking advantage the gauge invariance of the electromagnetic interaction and selecting an appropriate gauge, it has been shown that the Dirac oscillator interaction occurs as if it were produced by an infinite sphere carrying a uniform charge-density distribution, resulting in a linearly growing electric field \cite{bentez1990solution}. It has also been suggested that the Dirac oscillator interaction can be regarded as an alternative confinement potential for heavy quarks in quantum chromodynamics \cite{moreno1989covariance}. Hence, we can think that a well-established mathematical model for tight-binding fermion-fermion systems may be useful to better understand the physics behind such systems \cite{moshinsky1993barut,franco2013first,kulikov2007relativistic}.

\qquad The Dirac oscillator interaction has been studied in many areas of physics such as in mathematical physics \cite{zarrinkamar2010dirac,bermudez2008chirality,de2006relating,alhaidari2006dirac,quesne2005dirac,ho2004quasi,villalba1994exact}, nuclear (or subnuclear) physics \cite{munarriz2012spectroscopy,wang2012zitterbewegung,romera2011revivals} (also see \cite{grineviciute2012relativistic,faessler2005description}), quantum optics \cite{bermudez2008chirality,dodonov2002nonclassical,bermudez2007mesoscopic,bermudez2007exact,dutta2013pseudo} and molecular dynamics \cite{franco2013first}. One of the most important aspects of this system is that it has found many application areas in various branches of physics. The Dirac oscillator system one of the most useful tools in mathematical physics due to the fact it is an exactly solvable model. Thus, a great majority of the studies based on the determination of the influence of spacetime topology on the relativistic dynamics of the mentioned quantum systems are based on this useful system \cite{carvalho2011dirac,bakke2013interaction,salazar2019algebraic,bueno2014quantum,oliveira2019topological,hosseinpour2019dirac,bakke2013rotating}.  The Dirac oscillator has been applied in many other physical context such as graphene \cite{sadurni2011dirac} and it has been used to determine the quantum gravity effects \cite{benzair2017propagator,stetsko20191+}. However, it is worth to mention that the Dirac oscillator is normally utilized in the context of many-body theory \cite{moshinsky1993barut}.

\qquad A phenomenological way to describe the dynamics of many-body systems is to use one-time equations including free Hamiltonians for each particle plus interparticle interaction potentials \cite{kemmer1937,fermi1949}. Although, the history of such equations goes long way back \cite{van1997,giachetti2019}, one can see that there are still different relativistic two-body Dirac equations in use today in the literature \cite{van1997,giachetti2019,breit1929,salpeter1951,barut1985derivation,guvendi2019exact}. In this study, we deal with a fully-covariant two-body Dirac equation \cite{moshinsky1993barut,guvendi2019exact,guvendi2020interacting,guvendi2020relativistic,barut1985,barut1987a,barut1987b} that derived from quantum electrodynamics with the help of action principle \cite{barut1985derivation}. This equation includes the most general electric and magnetic potentials and leads a $16\times16$ matrix equation in $4$-dimensions \cite{barut1985,barut1987a,barut1987b}. Therefore, in $4$-dimensions, the separation of the angular and radial variables requires to be applied group theoretical methods \cite{barut1985,barut1987a,barut1987b,aydin1988relativistic}. It has been announced that the well-known energy spectrum for one-electron atoms or fermion-antifermion systems can be obtained via perturbative solution of the obtained radial equations \cite{moshinsky1993barut,barut1985,barut1987a,barut1987b}. Recently, it has been shown that this fully-covariant two-body Dirac equation can be exactly solvable for both low dimensional systems \cite{guvendi2019exact} and for systems that have dynamical symmetries \cite{guvendi2020relativistic,guvendi2020interacting}, without considering any group theoretical method. This fully covariant two-body Dirac equation can also be used to determine the relativistic dynamics of any fermion-fermion system in curved spacetimes \cite{guvendi2020interacting}. This equation has been used to determine the effect of gravitational field produced by cosmic strings on the energy of unstable composite systems such as positronium and the results, in principle, have shown that an ortho-positronium system have a potential to prove the existence of cosmic strings in the universe \cite{guvendi2020interacting}.

\qquad In this present manuscript, we introduce a unique presentation to determine the relativistic Landau levels of a fermion-antifermion pair interacting through Dirac oscillator interaction. In order to acquire this, we can write the fully-covariant two-body Dirac equation for arbitrary two Dirac oscillators interacting with an external uniform magnetic field. The system that under scrutiny has translational symmetry. Therefore, we can investigate the relativistic dynamics of such a composite system in $3$-dimensional flat Minkowski spacetime background by choosing the interaction of each particle with the external uniform magnetic field in the symmetric gauge. We can separate, covariantly, center of mass motion coordinates and relative motion coordinates and then arrive at a $4\times4$ dimensional matrix equation, resulting a set of coupled equations. For a static and spinless composite system formed by a fermion-antifermion pair interacting through Dirac oscillator interaction in the presence of external uniform magnetic field, the coupled equations lead to an exactly solvable $2^{nd}$ order wave equation. As a result, we can obtain a non-perturbative spectrum (in closed-form) in energy domain. The obtained energy spectrum shows that the system we deal with behaves like a relativistic quantum oscillator carrying total rest mass of such a composite system. Among the previously announced studies in the literature, one can find a few study that concentrate on the relativistic dynamics of fermion-fermion systems with Dirac oscillator interaction (see \cite{moshinsky1993barut,moshinsky1991anomalous,moshinsky1995supersymmetry,bednar1997connection}). We hope that our exact results can contribute to fill this gap in the literature. Moreover, we think that our results can give insight into the quark-antiquark systems.

\section{Mathematical Model}
\label{sec:1}
\qquad We suggest that the relativistic dynamics of a fermion-antifermion pair interacting through Dirac oscillator interaction in the presence of an external uniform magnetic field can be investigated, without considering any charge-charge interaction, by using the corresponding form of the following fully-covariant two-body Dirac equation \cite{moshinsky1993barut,guvendi2019exact,guvendi2020interacting,guvendi2020relativistic,barut1985,barut1987a,barut1987b},
\begin{eqnarray}
&\left\lbrace H^{\left(1\right)}\otimes\gamma^{0^{\left(2\right)}}+\gamma^{0^{\left(1\right)}}\otimes H^{\left(2\right)}  \right\rbrace \chi\left(\textbf{x}_{1},\textbf{x}_{2}\right)=0,\nonumber\\
&\resizebox{.90\hsize}{!}{$H^{\left(1\right)}=\left[\gamma^{\eta^{\left(1\right)}}\pi^{\left(1\right)}_{\eta}+ib_{1}\textbf{I}_{2}\right],\quad H^{\left(2\right)}=\left[\gamma^{\eta^{\left(2\right)}}\pi^{\left(2\right)}_{\eta}+ib_{2}\textbf{I}_{2}\right]$},\nonumber\\
&\resizebox{0.45\textwidth}{!}{$\pi^{\left(1\right)}_{\eta}=\left(\partial^{\left(1\right)}_{\eta}+i\frac{e_{1}A^{\left(1\right)}_{\eta}}{\hbar c}-\Gamma_{\eta}^{\left(1\right)}\right),\quad \pi^{\left(2\right)}_{\eta}=\left(\partial^{\left(2\right)}_{\eta}+i\frac{e_{2}A^{\left(2\right)}_{\eta}}{\hbar c}-\Gamma_{\eta}^{\left(2\right)}\right)$},\nonumber\\
&b_{1}=\frac{m_{1}c}{\hbar},\quad b_{2}=\frac{m_{2}c}{\hbar},\quad \left(\eta=0,1,2.\right),\nonumber\\
& \partial_{\eta}=(\partial_{0}, \partial_{1}, \partial_{2}.),\label{Eq1}
\end{eqnarray}
here, the superscripts $(1)$ and $(2)$ refer to the first fermion with mass $m_{1}$ and the second fermion with mass $m_{2}$, respectively, $\textbf{I}_{2}$ is $2$-dimensional unit matrix, $\Psi \left( \mathbf{x}_{1},\mathbf{x}_{2}\right)$ is the bi-local composite field that is constructed by Kronocker production $(\otimes)$ of arbitrary massive two Dirac fields as in the following,
\begin{eqnarray}
\chi \left( \mathbf{x}_{1},\mathbf{x}_{2}\right) =\chi_{1} \left( \mathbf{x}_{1}\right)\otimes \chi_{2} \left(\mathbf{x}_{2}\right),\label{Eq2}
\end{eqnarray}
$\hbar$ is the reduced Planck constant, $c$ is the speed of light, $e_{1}$ and $e_{2}$ represent the charges of these fermions, $\Gamma_{\eta}$ and $A_{\eta }$ stand for the spinor connections and vector potentials, respectively. Here, it is important to notice that, the $\gamma^{0}$ mean $\gamma^{\eta}\lambda_{\eta}$ in every where in Eq. (\ref{Eq1}) and the $\lambda_{\eta}$ is a timelike vector $\lambda_{\eta}=(1,0,0)$ \cite{barut1986center}. This equation includes spin algebra spanned by Kronocker productions of the Dirac matrices. More details about this equation can be found in \cite{barut1985derivation}. Thanks to the dynamic symmetry of the system that under scrutiny we can study in $3$-dimensional flat Minkowski spacetime background. In terms of cartesian coordinates, this spacetime background is represented by the following line element \cite{guvendi2019exact,guvendi2020relativistic},
\begin{eqnarray}
ds^{2}=c^{2}dt^{2}-(dx^{2}+dy^{2}).\label{Eq3}
\end{eqnarray}
For the system that under scrutiny in this spacetime background, it is clear that the spinor connections in Eq. (\ref{Eq1}) does not effect the dynamics, since they vanish (see \cite{sucu2007}). In the presence of the external uniform magnetic field, the vector potentials in Eq. (\ref{Eq1}) can be chosen in the symmetric gauge as follows \cite{guvendi2020relativistic},
\begin{eqnarray}
&A^{\left(1\right)}_{t}=0,\quad A^{\left(1\right)}_{x}=-\frac{B_{0}y_{1}}{2}, \quad A^{\left(1\right)}_{y}=\frac{B_{0}x_{1}}{2},\nonumber\\
&A^{\left(2\right)}_{t}=0,\quad A^{\left(2\right)}_{x}=-\frac{B_{0}y_{2}}{2}, \quad A^{\left(2\right)}_{y}=\frac{B_{0}x_{2}}{2},\label{Eq4}
\end{eqnarray}
in which the $B_{0}$ relates with the strength of external uniform magnetic field and also the written spatial coordinate pairs($x_{v},y_{v} \left(v=1,2.\right)$) correspond to the spatial coordinates of the particles in the spacetime background given in Eq. (\ref{Eq3}). According to the signature $(+,-,-)$ in Eq. (\ref{Eq3})
we can choose the Dirac matrices as in the following \cite{guvendi2019exact,guvendi2020relativistic},
\begin{eqnarray}
&\gamma^{t^{\left(1,2\right)}}=\sigma^{z},\quad \gamma^{x^{\left(1,2\right)}}=i\sigma^{x},\quad \gamma^{y^{\left(1,2\right)}}=i\sigma^{y},\nonumber\\
&\sigma^{x}=\left(
\begin{array}{cc}
0 & 1 \\
1  &0\\
\end{array}
\right),\quad \sigma^{y}=\left(
\begin{array}{cc}
0 & -i \\
i  &0\\
\end{array}
\right),\nonumber\\
&\quad \sigma^{z}=\left(
\begin{array}{cc}
1 & 0 \\
0  &-1\\
\end{array}
\right),\label{Eq5}
\end{eqnarray}
where, $\sigma^{x}$, $\sigma^{y}$ and $\sigma^{z}$ are the Pauli spin matrices. Also, we can introduce two dimensional two Dirac oscillators via the non-minimal substitution terms as follows \cite{ito1967example,cook1971relativistic,moshinsky1989dirac},
\begin{eqnarray}
&\resizebox{.90\hsize}{!}{$\partial_{x}^{\left(1\right)}\rightarrow\partial_{x}^{\left(1\right)}+\frac{m_{1}\omega_{1}}{\hbar}\gamma^{0^{\left(1\right)}}x_{1},\quad \partial_{y}^{\left(1\right)}\rightarrow\partial_{y}^{\left(1\right)}+\frac{m_{1}\omega_{1}}{\hbar}\gamma^{0^{\left(1\right)}}y_{1}$},\nonumber\\
&\resizebox{.90\hsize}{!}{$\partial_{x}^{\left(2\right)}\rightarrow\partial_{x}^{\left(2\right)}+\frac{m_{2}\omega_{2}}{\hbar}\gamma^{0^{\left(2\right)}}x_{2},\quad \partial_{y}^{\left(2\right)}\rightarrow\partial_{y}^{\left(2\right)}+\frac{m_{2}\omega_{2}}{\hbar}\gamma^{0^{\left(2\right)}}y_{2}$},\label{Eq5b}
\end{eqnarray}
in which $\omega_{1}$ is the frequency of the oscillator with mass $m_{1}$ and $\omega_{2}$ is the frequency of the oscillator with mass $m_{2}$. Now, one can separate the center of mass motion coordinates and relative motion coordinates via the following expressions \cite{guvendi2019exact},
\begin{eqnarray}
&R_{\eta}=\frac{1}{M}\left(m_{1}x_{\eta}^{\left(1\right)}+m_{2}x_{\eta}^{\left(2\right)}\right),\quad r_{\eta}=x_{\eta}^{\left(1\right)}-x_{\eta}^{\left(2\right)},\nonumber\\
&x_{\eta}^{\left(1\right)}=\frac{m_{2}}{M}r_{\eta}+R_{\eta},\quad x_{\eta}^{\left(2\right)}=-\frac{m_{1}}{M}r_{\eta}+R_{\eta},\nonumber\\
&\partial_{x_{\eta}}^{\left(1\right)}=\partial_{r_{\eta}}+\frac{m_{1}}{M}\partial_{R_{\eta}},\quad \partial_{x_{\eta}}^{\left(2\right)}=-\partial_{r_{\eta}}+\frac{m_{2}}{M}\partial_{R_{\eta}},\nonumber\\
&\partial_{x_{\eta}}^{\left(1\right)}+\partial_{x_{\eta}}^{\left(2\right)}=\partial_{R_{\eta}}.\label{Eq6}
\end{eqnarray}

\section{Coupled equations for the components of bi-spinor}
\label{sec:2}

\qquad Provided that the interaction is time-independent and the momentum of the center of mass is a constant of motion one can define the spinor $\chi$ as follows,
\begin{eqnarray}
&\chi \left(R_{0},\textbf{R},\textbf{r}\right)=e^{-iwt}e^{i\textbf{k}.\textbf{R}} \psi\left(\textbf{r}\right),\nonumber\\
&\psi\left(\textbf{r}\right)=\left(
\begin{array}{c}
\varphi_{1}\left(\textbf{r}\right)  \\
\varphi_{2}\left(\textbf{r}\right)  \\
\varphi_{3}\left(\textbf{r}\right)  \\
\varphi_{4}\left(\textbf{r}\right)
\end{array}%
\right),\label{Eq6b}
\end{eqnarray}
here, $w$ is total frequency that determined according to the proper time of the system ($R_{0}$) \cite{barut1985derivation}, $\textbf{k}$ relates with the center of mass momentum as $\hbar\textbf{k}$, $\textbf{R}$ is spatial position vector of the center of mass and the $\textbf{r}$ is spatial relative vector between the particles. Assuming the center of mass is located at the origin of the spacetime background and does not carry momentum ($\hbar\textbf{k}=0$) for oppositely charged ($e_{1}=e,e_{2}=-e$)\footnote{This case requires that $\omega_{1}=\omega,\omega_{2}=-\omega$ \cite{moshinsky1993barut}.} and equal massive ($m_{1}=m_{2}=m$) two Dirac oscillators in the presence of an external uniform magnetic field, we obtain the following matrix equation via the Eq. (\ref{Eq4}), Eq. (\ref{Eq5}), Eq. (\ref{Eq5b}), Eq. (\ref{Eq6}), Eq. (\ref{Eq6b}) and Eq. (\ref{Eq1}),
\begin{eqnarray}
&-\left(\gamma^{t}\otimes \gamma^{t}\right)i\varepsilon\psi+i \frac{mc}{\hbar}\left(\textbf{I}_{2}\otimes\gamma^{t}+\gamma^{t}\otimes \textbf{I}_{2}\right)\psi\nonumber\\
&+\left(\gamma^{x}\otimes \gamma^{t}-\gamma^{t}\otimes \gamma^{x}\right)\partial_{x}\psi\nonumber\\
&+\left(\gamma^{y}\otimes \gamma^{t}-\gamma^{t}\otimes \gamma^{y}\right)\partial_{y}\psi\nonumber\\
&+i\kappa x\left(\gamma^{y}\otimes \gamma^{t}+\gamma^{t}\otimes \gamma^{y}\right)-i\kappa y\left(\gamma^{x}\otimes \gamma^{t}+\gamma^{t}\otimes \gamma^{x}\right)\psi\nonumber\\
&+\xi x\left(\left(\gamma^{x}\gamma^{t}\right)\otimes \gamma^{t}+\gamma^{t}\otimes\left(\gamma^{x}\gamma^{t}\right)\right)\psi\nonumber\\
&+\xi y\left(\left(\gamma^{y}\gamma^{t}\right)\otimes \gamma^{t}+\gamma^{t}\otimes\left(\gamma^{y}\gamma^{t}\right)\right)\psi =0,\nonumber\\
&\varepsilon=\frac{w}{c},\quad \kappa=\frac{eB_{0}}{2\hbar c},\quad \xi= \frac{m\omega}{2\hbar},\label{Eq7}
\end{eqnarray}
in which $x$ and $y$ are the spatial coordinates of the relative motion. By multiplying the Eq. (\ref{Eq7}) with $\gamma^{t}\otimes\gamma^{t}$ from left (note that $\left(\gamma^{t}\otimes\gamma^{t}\right)^{2}$ gives $4\times4$ dimensional unit matrix), one can obtain the following $4\times4$ dimensional matrix equation,
\begin{eqnarray}
&\left(
\begin{array}{cccc}
\varepsilon-M &\partial_{-} &
-\partial_{-} & 0 \\
-\partial_{+}& \varepsilon & 0
& -\partial_{-} \\
\partial_{+} & 0 & \varepsilon &
\partial _{-} \\
0 & \partial_{+} & -\partial_{+} & \varepsilon+M
\end{array}
\right) \psi\left(\overrightarrow{r}\right)\nonumber\\
\nonumber\\
&\quad \quad \quad \quad \quad \quad +\left(\xi-\kappa\right)\left(
\begin{array}{cccc}
0 & r_{-} & r_{-}
& 0 \\
r_{+} & 0 & 0
&r_{-} \\
r_{+}  & 0 & 0 &
r_{-} \\
0 & r_{+} &r_{+}  & 0
\end{array}
\right)\psi\left(\overrightarrow{r}\right)=0,\nonumber\\
&M=\frac{2mc}{\hbar},\quad \partial _{\pm}=\partial _{x}\pm i\partial _{y},\quad r_{\pm}=x\pm iy. \label{Eq9}
\end{eqnarray}

\qquad In order to construct all possible spin eigen-states of the system that under scrutiny we can transform the spacetime background into the polar spacetime background with the help of the well-known expressions in follows \cite{guvendi2019exact,guvendi2020relativistic},
\begin{eqnarray*}
\partial _{\mp} = e^{\mp i\phi }\left(\mp
 \frac{i}{r}\partial _{\phi }+\partial _{r}\right),\quad r_{\mp}=re^{\mp i\phi },
\end{eqnarray*}
here, $\partial _{+}$ and $\partial _{-}$ stand for the spin raising and spin lowering operators, respectively, and $r$ is the radial distance between the particles. Only for the transformed bi-spinor,
\begin{eqnarray*}
\resizebox{0.97\hsize}{!}{$\psi\left(\overrightarrow{r}\right)\Longrightarrow\left(\varphi_{1}\left(r\right)e^{i\left(s-1\right)\phi},\varphi _{2}\left(r\right)e^{is\phi},\varphi_{3}\left(r\right)e^{is\phi},\varphi_{4}\left(r\right)e^{i\left(s+1\right)\phi}\right)^{\textbf{T}}$},
\end{eqnarray*}
we can obtain a set of coupled equations for the components,
\begin{eqnarray}
&\varepsilon \psi_{1}\left(r\right)-M\psi_{2}\left(r\right)+2\partial_{r}\psi_{3}\left(r\right)+2ur\psi_{4}\left(r\right)=0,\nonumber\\
&\varepsilon \psi_{2}\left(r\right)-M\psi_{1}\left(r\right)+\frac{2s}{r}\psi_{3}\left(r\right)=0,\nonumber\\
&\varepsilon \psi_{3}\left(r\right)+\frac{2s}{r}\psi_{2}\left(r\right)-2\left(\frac{1}{r}+\partial_{r}\right)
\psi_{1}\left(r\right)=0,\nonumber\\
&\varepsilon \psi_{4}\left(r\right)-2ur\psi_{1}\left(r\right)=0,\nonumber\\
&\psi_{1}\left(r\right)=\varphi_{1}\left(r\right)+\varphi_{4}\left(r\right),\quad \psi_{2}\left(r\right)=\varphi_{1}\left(r\right)-\varphi_{4}\left(r\right),\nonumber\\
&\psi_{3}\left(r\right)=\varphi_{2}\left(r\right)-\varphi_{3}\left(r\right),\quad \psi_{4}\left(r\right)=\varphi_{2}\left(r\right)+\varphi_{3}\left(r\right)\nonumber\\
&u=\left(\xi-\kappa\right),\label{Eq10}
\end{eqnarray}
here, $s$ represents to total spin of the composite system formed by a fermion-antifermion pair interacting through Dirac oscillator interaction in the presence of an external uniform magnetic field. Provided that the spins of these fermions are anti-symmetric with respect to each other, we can introduce a dimensionless independent variable that reads $z=u r^{2}$. In terms of the variable $z$, the equation system in Eq. (\ref{Eq10}) becomes as follows,
\begin{eqnarray}
&\resizebox{.95\hsize}{!}{$\varepsilon\psi_{1}\left(z\right)-M\psi_{2}\left(z\right)+4u\sqrt{\frac{z}{u}}\partial_{z}\psi_{3}\left(z\right)-2u\sqrt{\frac{z}{u}}\psi_{4}\left(z\right)=0$},\nonumber\\
\nonumber\\
&\varepsilon\psi_{2}\left(z\right)-M\psi_{1}\left(z\right)=0,\nonumber\\
\nonumber\\
&\varepsilon\psi_{3}\left(z\right)-\frac{2}{\sqrt{\frac{z}{u}}}\psi_{1}\left(z\right)-4u\sqrt{\frac{z}{u}}\partial_{z}\psi_{1}\left(z\right)=0,\nonumber\\
\nonumber\\
&\varepsilon \psi_{4}\left(z\right)-2u\sqrt{\frac{z}{u}}\psi_{1}\left(z\right)=0.\label{Eq11}
\end{eqnarray}
One can see that it is not difficult to obtain an exact solution of this set of equations written in the most symmetric form. We will obtain that this solution gives relativistic Landau levels of a spinless composite system consisting of a fermion-antifermion pair interacting through Dirac oscillator interaction, without requires any approximation.

\begin{figure}
\resizebox{.90\hsize}{!}{$\includegraphics{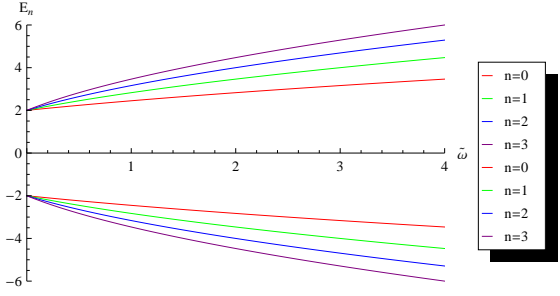}$}
\caption{The dependence of the total energy on the $\overline{\omega}$. Here, $m=c=\hbar=1$ and $0\leq\left(\omega-\frac{\omega_{c}}{2}\right)\leq1$.}
\label{fig:1}
\end{figure}

\section{Non-perturbative energy spectrum}
\label{sec:3}

\qquad Solving the equation system in Eq. (\ref{Eq11}) in favour of $\psi_{1}\left(z\right)$ one can acquire the following $2^{nd}$ order wave equation,
\begin{eqnarray*}
\resizebox{.98\hsize}{!}{$\partial^{2}_{z}\psi_{1}\left(z\right)+\frac{1}{z}\partial_{z}\psi_{1}\left(z\right)+\left(\frac{\varepsilon^{2}-M^{2}}{16uz}-\frac{1}{4 z^{2}}-\frac{1}{4}\right)\psi_{1}\left(z\right)=0$}.\label{Eq12}
\end{eqnarray*}
This wave equation can be reduced to the well-known shape of the Whittaker differential equation via the ansatz that reads $\psi_{1}\left(z\right)=\frac{1}{\sqrt{z}}\varsigma\left(z\right) $,
\begin{eqnarray}
\partial^{2}_{z}\varsigma\left(z\right)+\left(\frac{\varepsilon^{2}-M^{2}}{16uz}-\frac{1}{4}\right)\varsigma\left(z\right)=0.\label{Eq13}
\end{eqnarray}
Solution function of Eq. (\ref{Eq13}) is obtained as follows \cite{guvendi2020relativistic,book:874771,dernek2018relativistic},
\begin{eqnarray*}
\varsigma\left(z\right)=NW_{\varpi,\tau}(z ),\quad \varpi= \frac{\varepsilon^{2}-M^{2}}{16uz},\quad \tau=\frac{1}{2},
\end{eqnarray*}
in which $N$ is the normalization constant. However, this solution function can only be reduced to polynomial of degree $n\geq0$ with respect to the variable $z$ provided that the following condition holds \cite{guvendi2020relativistic,book:874771,dernek2018relativistic},
\begin{eqnarray}
\frac{1}{2}+\tau-\varpi=-n,\label{Eq14}
\end{eqnarray}
here, $n$ is the principal quantum number (non-negative integer). Now, one can immediately obtain a non perturbative spectrum in energy domain with the help of the expressions given in Eq. (\ref{Eq13}) and Eq. (\ref{Eq14}), since Eq. (\ref{Eq14}) leads to the quantization condition for the formation of the composite system that under scrutiny,
\begin{eqnarray}
&E_{n}=\pm 2mc^{2}\sqrt{1+\frac{\overline{\omega}\hbar}{2mc^{2}}\left(n+1\right)},\nonumber\\
& \overline{\omega}= 4\left(\omega-\frac{\omega_{c}}{2}\right),\quad \omega_{c}=\frac{|e|B_{0}}{m c},\label{Eq15}
\end{eqnarray}
here, $\omega_{c}$ stands for the cyclotron frequency \cite{guvendi2020relativistic,bermudez2007mesoscopic}. The Eq. (\ref{Eq15}) gives the relativistic Landau levels of a static and spinless composite system formed by a fermion-antifermion pair interacting through Dirac oscillator interaction. This non-perturbative energy spectrum shows that the system that under scrutiny behaves like a single relativistic quantum oscillator carrying total rest mass of the composite system. In Eq. (\ref{Eq15}), one can see that the second term in square root does not vanish even for the ground state of such a system. We can also see that the oscillation of this composite system will stop if and only if $\omega=\omega_{c}/2$ holds for $e<0$ and the total energy of the system becomes equal to the total rest mass energy of the particles when $\omega=\omega_{c}/2$. Also, in terms of the variable $z$, we can easily obtain the components of the bi-spinor as in the following,

\begin{eqnarray}
\centering
&\psi_{1}\left(z\right)=N\frac{W_{\varpi ,\tau}(z )}{\sqrt{z}},\nonumber\\
&\psi_{2}\left(z\right)=N\frac{M}{\varepsilon \sqrt{z}}W_{\varpi ,\tau}(z ),\nonumber\\
&\psi_{3}\left(z\right)=N\left[\frac{2}{\varepsilon \sqrt{\frac{1}{u}}}+\frac{4\sqrt{uz}}{\varepsilon}\left(\frac{1}{2}-\frac{\varpi}{z}\right)\right]W_{\varpi ,\tau}(z )\nonumber\\
&-N\frac{4\sqrt{u}}{\varepsilon z}W_{\varpi+1 ,\tau}(z ),\nonumber\\
&\psi_{4}\left(z\right)=-N\frac{2\sqrt{u}}{\varepsilon}W_{\varpi ,\tau}(z ),\label{Eq16}
\end{eqnarray}
which satisfy the coupled equations in Eq. (\ref{Eq11}), of course.

\section{Results and Discussion}
\label{sec:4}

\qquad In this paper, we introduce a unique model for a composite system formed by a fermion-antifermion pair interacting through Dirac oscillator interaction in the presence of an external uniform magnetic field. This model based on an exact solution of the corresponding form of a fully-covariant two body Dirac equation (one-time) including spin algebra spanned by Kronocker productions of the Dirac matrices.  The translational symmetry of the interested system allows to study in $2+1$ dimensional flat Minkowski spacetime background. We transform the spacetime background into the polar spacetime background so that we can exploit the angular symmetry. This transformation provides to obtain an exactly solvable $2^{nd}$ order wave equation for such a composite system (see also \cite{moshinsky1993barut,moshinsky1991anomalous}). We solve this wave equation and arrive at a non-perturbative spectrum (in closed-form) in energy domain. The obtained spectrum (Eq. (\ref{Eq15})) shows that, the total energy ($E_{n}$) of the system closes to total rest mass energy ($2mc^{2}$) of this composite system (static) when both of the external magnetic field and oscillator frequency close to zero. It is clear that in Eq. (\ref{Eq15}), the term associated with the both external magnetic field and oscillator frequencies does not vanish for such a composite system in the ground state ($n=0$). We can also see that, the obtained total energy value closes to the total rest mass energy of the system when $\omega\approx \frac{\omega_{c}}{2}$ and the oscillation of this composite system will stop if and only if $\omega = \frac{\omega_{c}}{2}$ is satisfied. The obtained energy spectrum in Eq. (\ref{Eq15}) shows that the system that under scrutiny behaves like a single relativistic quantum oscillator carrying total rest mass of the composite system in the presence of external uniform magnetic field. The dependence of the total energy on the $\overline{\omega}$ can be seen in Fig. $1$. As a result, our unique presentation in this manuscript provides to obtain the relativistic Landau levels of a spinless composite system consisting of a fermion-antifermion pair interacting through Dirac oscillator interaction. We think that the obtained non-perturbative results in this paper can give info about the quark-antiquark systems and thus mass formula for mesons (see also \cite{moshinsky1993barut}).


\begin{acknowledgements}
The author thanks Özge SAKARYA ÇINKI for style suggestions.
\end{acknowledgements}



\end{document}